\documentclass[letterpaper,11pt]{article}
\usepackage{amsthm}
\usepackage{amsfonts}
\usepackage{amsmath}
\usepackage{amssymb}
\usepackage{graphicx}
\usepackage{epstopdf}
\usepackage{lineno}
\usepackage{setspace} 
\usepackage{verbatim}
\usepackage{hyperref}
\usepackage{authblk}
\usepackage{color}
\usepackage{textcomp}

\newtheorem*{theorem*}{Theorem}

\newenvironment{enumerate*}{
\begin{enumerate}
  \setlength{\itemsep}{0pt}
  \setlength{\parskip}{0pt}
  \setlength{\parsep}{0pt}
}{\end{enumerate}}

\newenvironment{itemize*}{
\begin{itemize}
  \setlength{\itemsep}{5pt}
  \setlength{\parskip}{0pt}
  \setlength{\parsep}{0pt}
}{\end{itemize}}

\title{ Algorithm engineering for a quantum annealing platform}
\author[1]{Andrew D.~King\thanks{Corresponding author.}}
\author[1,2]{Catherine C.~McGeoch}
\affil[1]{D-Wave Systems, Burnaby,  BC \{aking, cmcgeoch\}@dwavesys.com }
\affil[2]{Department of Computer Science,  Amherst College, Amherst, MA}

\begin{document}
\maketitle

Recent advances bring within reach the viability of solving combinatorial problems using a quantum annealing algorithm implemented on a purpose-built platform that exploits quantum properties.  However, the question of how to tune the algorithm for most effective use in this framework  is not well understood.  In this paper we describe some operational  parameters that drive performance, discuss approaches for mitigating sources of error, and present experimental results from a D-Wave Two quantum annealing processor.  

\thispagestyle{empty}

\newpage
\setcounter{page}{1} 

\section{Introduction}

In the last three decades researchers in algorithm engineering have
 identified many strategies for bridging the gap between abstract algorithm and concrete implementation  to yield practical performance improvements;  see \cite{McGeoch2012} or \cite{Mueller-Hannemann2010} for an overview. 
In this paper  we apply this conceptual framework in a novel context,  to improve performance 
of a {\em quantum annealing} algorithm implemented on a purpose-built platform.  
Quantum annealing (QA) is a heuristic method 
for solving combinatorial optimization problems,  similar to simulated annealing.   The platform is a D-Wave Two\footnote{D-Wave,  D-Wave Two, and Vesuvius are trademarks of D-Wave Systems Inc.} system, which exploits quantum properties to solve instances of the NP-hard Ising Minimization Problem (IMP).  

Several research groups have reported on experimental work to understand performance of D-Wave systems; see for example  \cite{bian2014discrete}, \cite{boixo2014evidence}, \cite{henaqc2014}, \cite{McGeoch2013}, \cite{Pudenz2014a}  \cite{Roennow2014}, and \cite{Venturelli2014}.  Building on this experience we describe an emerging performance model that helps to distinguish the {algorithm}  from its  {realization} on a physical platform.  Using this model we present a collection of strategies for improving computation times in practice.  Our discussion exposes similarities as well as differences in 
algorithm engineering approaches to  quantum versus classical computation.  

The remainder of this section presents a quick overview of  the quantum annealing algorithm and its
realization in D-Wave hardware.  Section 2 surveys the main factors that drive performance.  Section 3 
presents our strategies together with experimental results to study their efficacy.  Section 4 presents a few concluding remarks.  

\paragraph{The native problem}  
An input instance to the {\em Ising Minimization Problem} (IMP) is described by a {\em Hamiltonian} $(h,J)$ containing a vector of {\em local fields} $h \in \mathbb R^n$  and a matrix of {\em couplings}  
$J \in \mathbb R^{n \times n}$ (usually upper-triangular).   We may consider weights $h_v$ and nonzero 
 $J_{uv}$ to be assigned to vertices and edges of a graph $G=(V,E)$. 
The problem is to  find an assignment of {\em spin} values $s\in \{-1,1\}^n$ (i.e.\ a {\em spin configuration} or {\em spin state}) that minimizes the function
\begin{eqnarray}
E(s) = E(h,J,s) := \sum_{v\in V(G)}h_v s_v + \sum_{uv\in E(G)} J_{uv} s_us_v.
\end{eqnarray}  
This problem has origins in statistical physics,  where $E(s)$  defines the {\em energy} of a 
given {spin state} $s$.  A {\em ground state} has minimum energy.  A non-ground 
state is called an {\em excited state};  a  {\em first excited
state} has the lowest energy among exited states.  
  Notice how the signs of $(h,J)$ affect this  function:  
a term  with $J_{uv} < 0$, called a {\em ferromagnetic coupling},  is  minimized when $s_u = s_v$; an {\em antiferromagnetic coupling} term with $J_{uv} > 0$ is minimized when $s_u \neq s_v$.   The problem is NP-hard when $G$ is nonplanar \cite{Istrail2000}.  

\paragraph{Quantum annealing}            
While a classical bit takes discrete values 0 or 1,  a quantum bit (qubit) is capable of {\em superposition}, which means that is simultaneously in both states; thus a register of $N$ qubits can represent all $2^N$ possible states simultaneously.  When a qubit is read,  its superposition state ``collapses'' probabilistically to a classical state,  which we interpret as a spin $-1$ or $ +1$.  

Qubits act as  particles in a quantum-mechanical system that evolves under forces described by a time-dependent 
Hamiltonian ${\mathcal H}(t)$.   For a given Hamiltonian they naturally seek their ground state just as water seeks the lowest point in a landscape.  Since superposition is represented not by a single state but by a probability mass,  we can think of it as moving through hills in a porus landscape -- this is sometimes called {\em tunneling}.  A quantum annealing algorithm exploits this property to perform an analog computation defined by the following components.  

\begin{enumerate*}
\item[(C1)] The  {\em initial Hamiltonian}  ${\cal H}_I$ puts each qubit into superposition whereby spins are independent and equiprobable.   

\item[(C2)] The {\em problem Hamiltonian}  ${\cal H_P}=(h,J)$ matches the objective function (1) so that a ground state corresponds to an optimal solution to the problem.   

\item[(C3)] The {\em  path functions}  $A(r), B(r)$ define a  transition from ${\cal H}_I$  to ${\cal H}_P$, where 
$A(r): 1 \rightarrow 0$ and $B(r): 0 \rightarrow 1$ as  $r: 0 \rightarrow 1$.  
Parameter $r$ controls the rate of change $r = r(t)$ (possibly speeding up or slowing down)  as time $t$ moves from start $t_0=0$ to finish $t_f$.  
\end{enumerate*} 
The entire  algorithm is defined by the time-dependent Hamiltonian:  
\begin{eqnarray}
{{\cal H}(t)} =  A(r) {\cal H}_I  + B(r) {\cal H}_P. 
\label{eq:ham} 
\end{eqnarray}    

 A QA algorithm can be simulated classically using many random states to model superposition:  (C3) is analogous to a simulated 
annealing schedule, except it modifies the problem landscape rather than a  traversal probability;  (C1) corresponds to choosing random initial states in a flat landscape;  and (C2) to the target solution.  See \cite{Das2005}, \cite{Kadowaki1998} or \cite{Martonak2002}.  
QA belongs to the {\em adiabatic}  model of quantum computation (AQC), which is  a polynomially-equivalent \cite{aharonov2007adiabatic,Farhi2001} alternative to the more 
familiar quantum gate model.  QA algorithms typically use problem Hamiltonians from a subclass of those in  the full AQC model;  thus QA computation is 
likely not universal, although the question is open (see \cite{McGeoch2014}).

\section{Hardware platform and cost models}

A D-Wave Two (DW2) platform contains a quantum annealing chip that physically realizes the algorithm in Equation \ref{eq:ham}.  
Qubits and the couplers connecting them are made of microscopic superconducting loops of niobium, which exhibit quantum properties at the processor's operating temperature, typically below 20mK. See \cite{bunyk2014architectural} for an overview.

The annealing  process is managed by a framework of analog control devices that relay signals between a conventional CPU and the qubits and couplers onboard the chip, in stages as follows.  
\begin{enumerate*}
\item {\em Programming. }  The weights $(h, J)$ are loaded onto the chip.  Elapsed time = $t_p$. 
\item {\em Annealing. } The algorithm in (2) is carried out.   Time = $t_f$.    
\item {\em Sampling.}   Qubit states are measured,  yielding a solution $s$. Time = $t_s$.  
\item {\em Resampling.}  Steps 2 and 3 are repeated to obtain some number $k$ of sampled solutions.    
\end{enumerate*}
Total computation time is therefore equal to  
\begin{eqnarray}
T = t_{p}  +  k (t_{f}  + t_{s}).
\end{eqnarray}

Component times vary from machine to machine;  the system used in our tests has operating parameters shown in Table \ref{tab:specs}.   Note that total time is dominated by what are essentially I/O costs;  successive processor models have generally shown reductions in these times and this trend is expected to continue.   Anneal time  $t_f$ can be set by the programmer;  the minimum setting $20\mu s$  is dictated by the system's ability to shape $A(r)$ and $B(r)$.  

\begin{table}
\begin{centering} 
\begin{tabular}{|rrrrrrr|} \hline 
name & qubits & couplers  & temperature &   $t_p$  &   $t_f$          & $t_s$  \\ \hline
V7   &  481   & 1306     & 14mK  &  30ms    &  20$\mu$s     &  116$\mu$s  \\   \hline
\end{tabular}\\~
\end{centering}
\caption{\small{Operating parameters for the processor used in our tests.}}\label{tab:specs}
\end{table}

\subsection{Analysis} 
In algorithm engineering we  can identify different levels of {\em instantiation} in a spectrum that includes the pencil-and paper algorithm,  an implementation in a high-level language,  and a sequence of machine instructions.  The definition of time performance (dominant cost vs.\ CPU time) and the set of strategies for reducing it  (asymptotics vs.\ low-level coding) depend  on the level being considered.   This framework applies to quantum as well as classical computation.  This subsection describes instantiation layers 
and cost  models for the quantum annealing algorithm realized on D-Wave platforms.  

\paragraph{Asymptotics of closed-system AQC} 
Abstract AQC algorithms have been developed for many computational problems;  see \cite{McGeoch2014} for examples.  
 For a given algorithm (a generalization of (2)) and  input of size $n$, 
let $\gamma = \gamma(n)$ denote the {\em minimum spectral gap},  the smallest difference between 
the energies of the ground state and the first excited state at any time $t: 0 \rightarrow t_f$.  
Under certain assumed conditions, if 
$t_f$ is above a threshold in $\Theta(poly(n)/\gamma^2)$, then the computation will almost surely finish in ground state. Setting  $t_f$ below the threshold increases the probability that a nonoptimal solution is returned.  
 Typically $\gamma$ is difficult to compute and bounds are known only for simple scenarios; some algorithm design strategies 
have been identified for ``growing the gap'' to reduce asymptotic computation times.  See  \cite{farhiaqc2014}, \cite{Farhi2001},  \cite{Jordan2006}, \cite{McGeoch2014}, or \cite{nishimori2014comparative} for more.  

\paragraph{Quantum computation in the real world} 
Asymptotic analysis assumes  that the algorithm runs in a {\em closed system} in perfect isolation from 
external sources of energy (thermal, electrical, magnetic, etc).  It is a matter of natural law, however, that  any physically-realized quantum computer runs  in 
an {\em open system} and suffers interference from the environmental ``energy bath.''    Environmental interference may reduce the probability of 
finishing in ground state -- in particular the theoretical annealing time threshold depends on both $\gamma^2$ and the ambient temperature \cite{albash2012quantum}, implying that colder is faster.  In practice, there is evidence that the thermal bath can increase the probability of success substantially \cite{Dickson2013}.

We use the terms AQC and QA to distinguish algorithms running in closed vs.\ open systems.  Some (exponential)  bounds on convergence times of classical QA algorithms are known \cite{Kadowaki1998, Roland2002};  these bounds are better than those of simulated annealing in some cases.   

\paragraph{Realization on D-Wave platforms}  
In addition to the above nonideality,  DW2 architecture imposes some restrictions on inputs:    
\begin{enumerate} 

\item The connection topology defines a {\em hardware graph} $G=(V,E)$, a subgraph of a  $\mathcal C_8$ {\em Chimera} graph \cite{bunyk2014architectural} containing 512 qubits.    An IMP instance defined on a general graph $G_0$ must be  {\em minor-embedded} onto $G$.  This requires $O(n^2)$ expansion in problem size 
\cite{Choi2008} in the worst case;  in practice we use a heuristic approach described in \cite{Cai2014}.  See Appendix A for more about Chimera graphs and minor-embeddings.  

\item  The elements of $h$ and $J$ must be in the real range $[-1,1]$.  This can be achieved by scaling 
general $h$ and $J$ by a positive constant factor $\alpha$.    

\item The weights $(h,J)$, specified as floats, are transmitted imperfectly by the analog control circuitry. As a result, they experience perturbations of various sorts, systematic (biased), random, persistent, and transient.\footnote{This is in contrast to the meme that Hamiltonian misspecification is due to calibration errors (cf.\ \cite{Roennow2014}). Calibration errors, which are systematic and relatively fixed, represent only a small component of ICE.}  The perturbations are collectively referred 
to as {\em intrinsic control error} (ICE).  
Because of ICE, the problem Hamiltonian solved by the chip may be slightly different from the problem Hamiltonian specified by the programmer. 
\end{enumerate}  

Putting all this together, total computation time in Equation (3) depends on the probability $\pi$ of observing a successful outcome (a ground state) in a single sample.  In theory,  $\pi$ depends on a threshold value for $t_f$, which is typically unknown in open-system computing.   Because of Hamiltonian misspecification we may prefer  $\pi < 1$ in order to  sample solutions near the 
(wrong)  ground state;  if $\pi$ is too small,  $k$ can be increased to improve the overall success probability.   Just as in classical computing,  there is a trade-off  between time and solution quality,  although very little is known about the nature of that trade-off.  

In what follows, we calculate the {\em empirical success probability} $\pi$ for a given input as the proportion of successful samples drawn among $k$ samples from the hardware, using various definitions of success in order to examine the relationship between computation time and solution quality.   We calculate the expected number of samples required to observe a successful outcome with probability at least $0.99$ (ST99): this is $k_{99} = \log ( 1- .99)/ \log (1-\pi)$.   Computation time is found by combining  $k_{99}$ with component times as in  (3).   

\subsubsection{ICE:  The error model}\label{sec:error} 
Our simplified model of ICE, which will be described more fully in a forthcoming paper, assumes that the problem Hamiltonian $(h,J)$ is perturbed by an error Hamiltonian $(\tilde h,\tilde J)$, where $\tilde h_u$ and $\tilde J_{uv}$ are independent Gaussians having mean 0 and standard deviations $\sigma_h$ and $\sigma_J$, which vary by chip (and generally decrease with new models).  For V7 (see Table \ref{tab:specs}) we have  $\sigma_h\approx .050$ and $\sigma_J\approx .035$.   These errors are relative to the nominal scale of $[-1, +1]$, which means that if $h$ and $J$ are scaled by $\alpha \in (0,1)$, relative errors are amplified by a factor of $1/\alpha$.\ \footnote{This holds for most sources of ICE, with the notable exception of {\em background susceptibility}, denoted $\chi$, which is reduced by a factor of $1/\alpha$ -- see \cite{Vinci2014}.  Background susceptibility is an instance-dependent, non-transient error that, in a more sophisticated error model, might be separated from Gaussian error.}

For a given spin configuration $s$ this shifts the effective energy from $E(s)$ by a Gaussian error $\tilde E(s)$ with mean 0 and standard deviation  $\sigma_E=(N\sigma_h^2  + M\sigma_J^2 )^{1/2}$ where $N$ and $M$ are the number of active qubits and couplers in the hardware graph.  
On a full size V7 problem ($N=481$)  we have  $\sigma_E=1.67$.  By the three-sigma rule, $|\tilde E(s)|<1.67$ and $|\tilde E(s)|<5.01$ for about 68 and 99.5 percent of spin configurations, respectively.  Although $\sigma_E$ scales as $\Theta(\sqrt n)$, the typical value of $\max_s\{|\tilde E(s)|\}$ scales linearly in $n$, and is near $14$ at full size.  

ICE imposes a practical limit on the precision of (scaled) weights that can be specified in successful computations.  For example, if  $h_u, J_{uv} \in [-1,+1]$, then two solutions $s$ and $s'$ with $E(s)<E(s')$ satisfy $E(s)\leq E(s')-2$, so it is relatively unlikely that $E(s)+\tilde E(s) > E(s')+\tilde E(s')$.  The difficulty occurs when energy levels differ by smaller amounts, which can happen when integer weights are scaled by $\alpha < 1$.

\begin{figure}
\begin{center}
\includegraphics[height=2.2in]{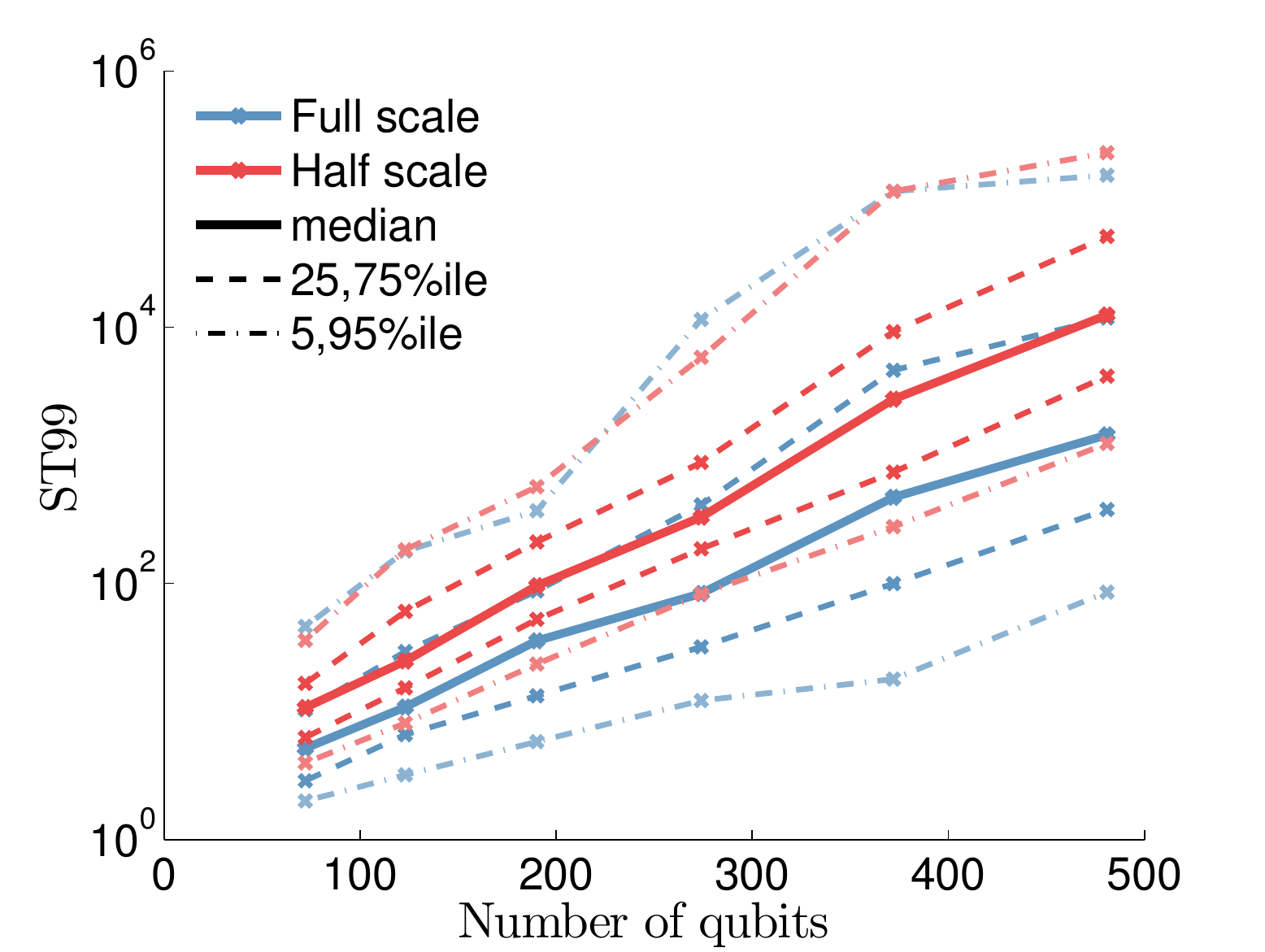}%
\includegraphics[height=2.2in]{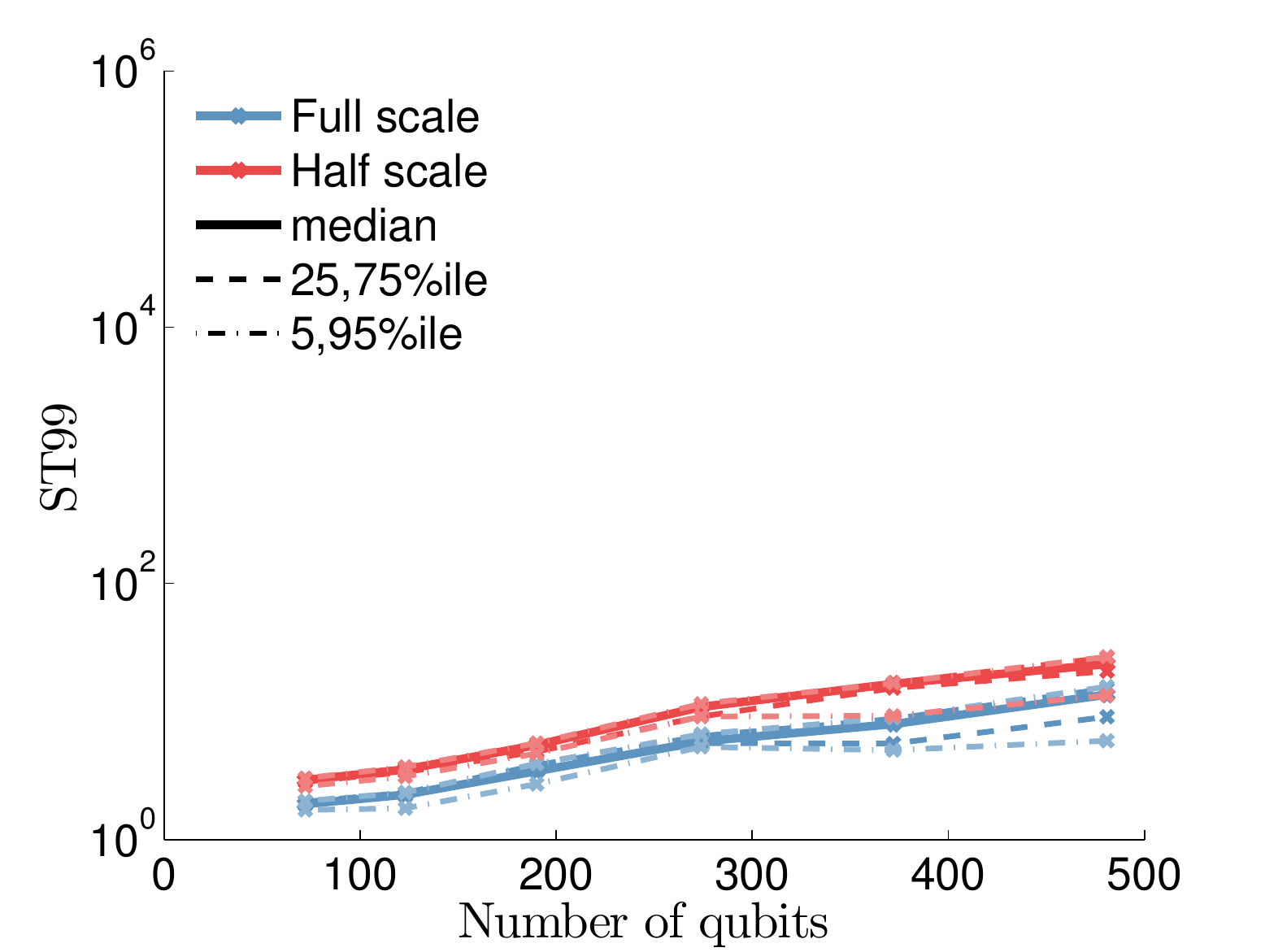}%
\end{center}   
\caption{\small{RAN3 problems solved at full scale (blue) and half scale (red).  Left panel shows ST99 for exact solution; right panel shows ST99 for solution within $\sigma_E \approx 1.67\sqrt{N/481}$ of ground state energy.  Lines show 5th, 25th, 50th, 75th, and 95th percentiles of 100 inputs for each size.}}
\label{fig.error-model} 
\end{figure} 

Figure \ref{fig.error-model} illustrates this effect using RAN3 instances (described in the next section) solved at full scale ($\alpha=1/3$) and half-scale ($\alpha=1/6$).  The left 
panel shows how reducing the problem scale increases ST99 roughly tenfold in the median case for largest problems when searching for an optimal solution.  
The right panel shows ST99 when the success condition is to find a solution within $\sigma_E$ of ground state.  In both scales,  computation times shrink by more than two orders of magnitude in nearly all percentiles.  This suggests that reductions in ICE on future chip models are likely to boost hardware performance significantly.  Analyzing performance with respect to the $\sigma_E$ error bound allows us to look beyond the effect of Hamiltonian misspecification, which is  detrimental to hardware success rates and may mask evidence of quantum speedup.

\section{Algorithm Engineering on D-Wave Two Platforms} 
In this section we consider strategies for mitigating ICE-related nonideality and small spectral gaps with the goal of increasing success probabilities and lowering computation times.   
 
D-Wave systems realize a specific QA algorithm in the sense that the components  ${\cal H}_I,  A(r), B(r)$ and $r(t)$ are set in 
firmware (see \cite{bunyk2014architectural} for details).   Here  we focus on parameters that can be controlled  by the programmer,  
namely  $(h, J)$, $t_f$, and $k$.  We also consider classical methods for pre-processing and error correction.  
We evaluate these strategies on the following instance classes, described more fully in Appendix \ref{appendix:testbed}.
\begin{itemize}

\item Random native instances (RAN$R$).   For each $(u,v) \in G$,  $J_{uv}$ is assigned a random nonzero integer in $\{-R \ldots +R\}$.  We set $h_u = 0$.
\footnote{ Katzgraber et al.\ \cite{Katzgraber2014} have shown that these instances are not suitable for investigating quantum speedup because the solution landscape has many global minima and no nonzero-temperature phase transition.  Consequently heuristic search algorithms act almost as random samplers, and there is no evolution of tall, thin barriers that would allow an open-system quantum annealer to exhibit an advantage through tunneling.  However, this class is suitable for looking at non-quantum effects such as ICE, as we do here.}

\item Frustrated loop instances (FL$R$) \cite{henaqc2014}.  These are constraint satisfaction problems whose entries 
of $J$ lie in $\{-R,\ldots,R\}$.  They are combinatorially more interesting than RAN$R$ instances  but do not require minor-embedding. 

\item Random cubic MAX-CUT instances (3MC).  These are MAX-CUT problems on random cubic graphs, which must be minor-embedded onto the V7 hardware graph.

\item Random not-all-equal 3-SAT instances (NAE).  These are randomly generated problems near the SAT/UNSAT phase transition,  filtered subject to having a unique solution (up to symmetry), and then minor-embedded onto the V7 hardware  graph.
\end{itemize}  
All experiments described here take random instances generated at sizes of up to $481$ qubits; the specifics of instances are given in Appendix A.  Unless otherwise specified,  $k_{99}$ is calculated from 1000 samples in  10 gauge transformations (next section), totaling 10,000 samples.  Optimal solutions are verified using an independent software solver.  In rare cases a sample will not contain an optimal solution, giving an empirical success probability of 0 and ST99 $= \infty$.  To simplify data analysis we look at ST99 for the 95th and lower percentiles of each input set;  missing percentile points in some graphs correspond to observations of $\pi=0$.     

\subsection{Gauge transformations}
Given instance $\mathcal H = (h,J)$, one can construct a modified instance $\mathcal H'=(h',J')$ by flipping the sign of some subdimension of the search space, as follows:  take a vector $\vec g\in\{-1,1\}^n$, set $h'_u=h_ug_u$ for each $u$, and set $J'_{uv}= J_{uv}g_ug_v$ for each coupler $uv$.    When solving $\mathcal H$ in hardware, we can divide the $k$ samples among $p$ instances $\mathcal H_1, \mathcal H_2,\ldots, \mathcal H_p$, where $\mathcal H_i$ is constructed from $\mathcal H$ by a random gauge transformation $\vec g^{(i)}$; we then apply the (idempotent) transformation to the hardware output to obtain a solution for $\mathcal H$.  Doing this mitigates the effects of some sources of ICE.  Gauge transformations are also described in \cite{boixo2014evidence} and \cite{Roennow2014}.

\begin{figure}
\begin{center}
\includegraphics[height=2.2in]{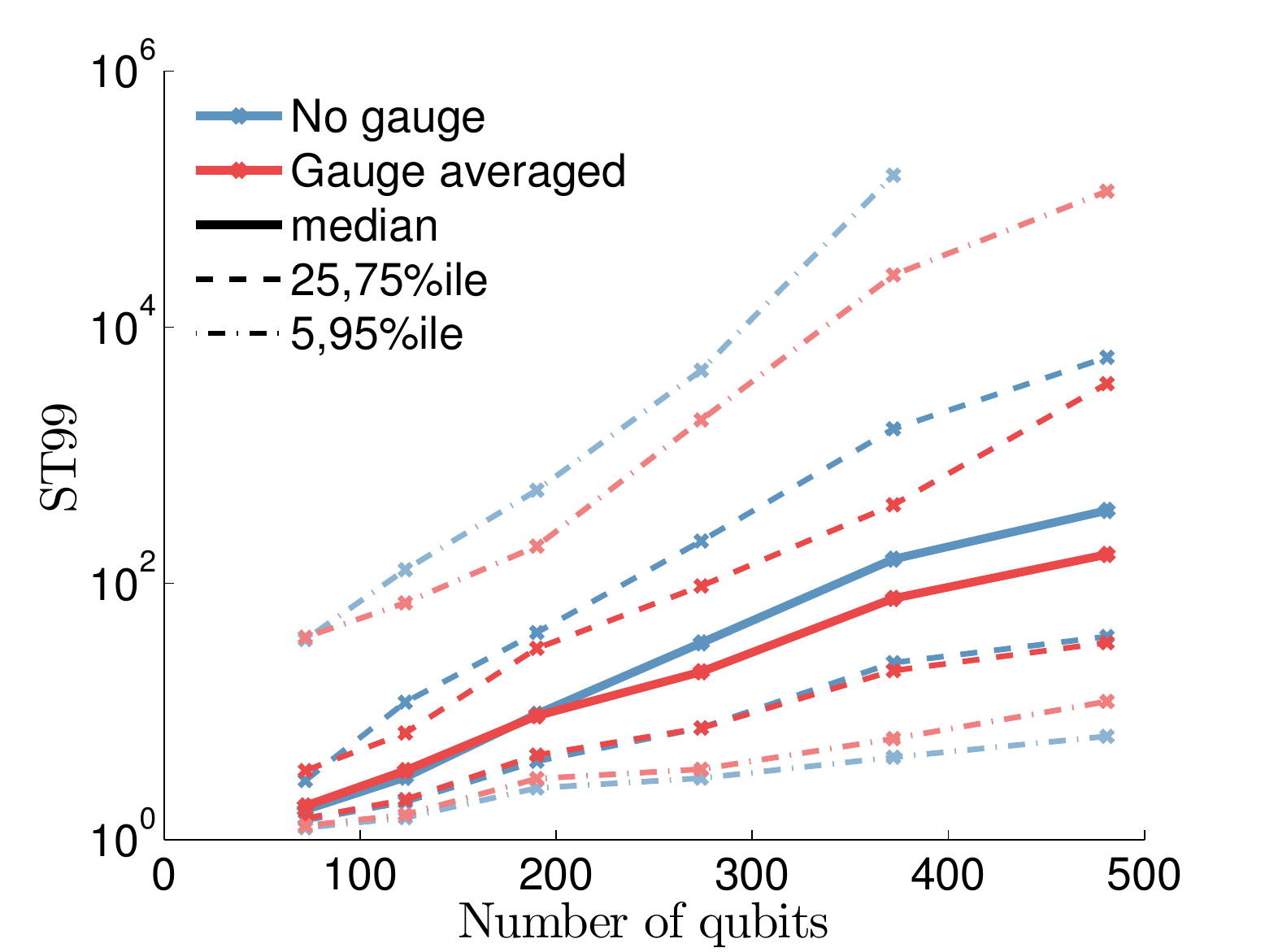}%
\includegraphics[height=2.2in]{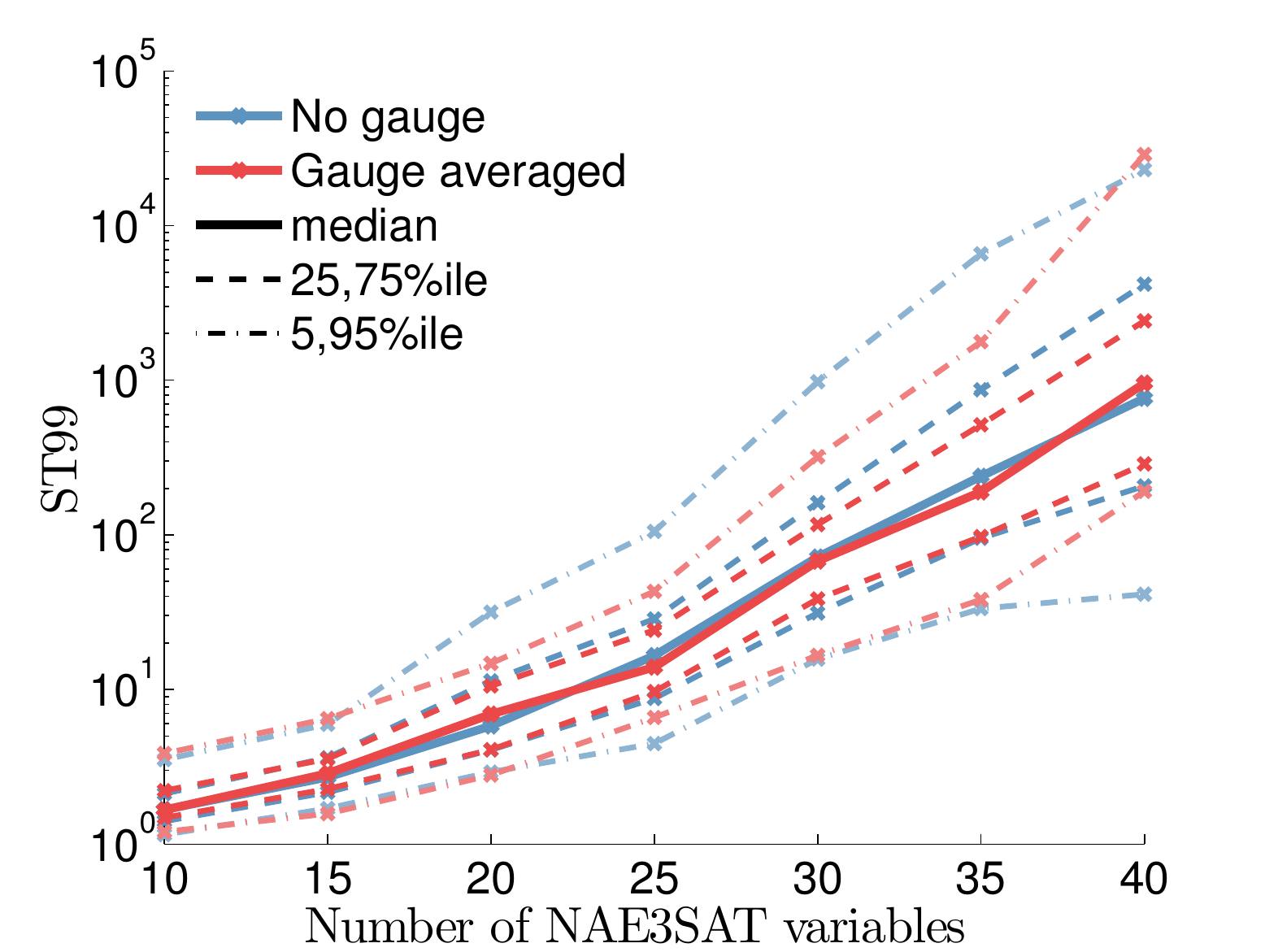}%
\end{center}   
\caption{\small{Effect of gauge transformations on RAN1 (left) and NAE (right) instances.   Times marked in blue take 10,000 samples with no gauge transformation; times in red take 1000 samples at each of  $p=10$ gauge transformations.  Percentile lines are shown for 100 and 50 inputs at each problem size, respectively. }}
\label{fig.gt}
\end{figure}

Figure \ref{fig.gt} shows the effect of applying $p=10$  gauge transformations on RAN1 instances (left) and NAE instances (right).   In both cases, gauge transformations help more on the most difficult problems (higher percentiles).   This is unsurprising, as difficult problems are typically more sensitive to perturbation by ICE.  Note that every $\mathcal H_i$ is a new instance which requires a programming step;  the current dominance of $t_p$ over $t_f+t_s$ means that it is rarely cost-effective to draw fewer than 1000 samples per gauge transformation.  However, this technique may yield more significant  performance improvements in applications other than optimization, such as  fair sampling of the solution space,  which is highly sensitive to Hamiltonian misspecification.  

\subsection{Optimal anneal times}  
Previous work \cite{boixo2014evidence,Roennow2014} has reported on experiments to find optimal settings of $t_f$ for RAN$R$ instances, concluding that for problem sizes $N \leq 512$  the lowest possible $t_f = 20\mu$s is longer than optimal. More recent work has found instances whose optimal anneal time on a DW2 processor is greater than $20\mu$s \cite{lidarcomm}.  Those studies consider anneal time in isolation,  so that the optimal time $t_f$ minimizes $k_{99}t_f$.   However, under the cost model in (3),  the optimal $t_k$ minimizes $k_{99}( t_k + t_s)$ so that a smaller increase in $\pi$ is sufficient to reduce total runtime in practice.  Also, by analogy to observations about simulated annealing in \cite{Venturelli2014},  we might expect  that longer anneal times are optimal for problem classes that are combinatorially interesting but relatively insensitive to misspecification (compared to RAN$R$ instances for large $R$).

\begin{figure}
\begin{center}
\includegraphics[height=2.1in]{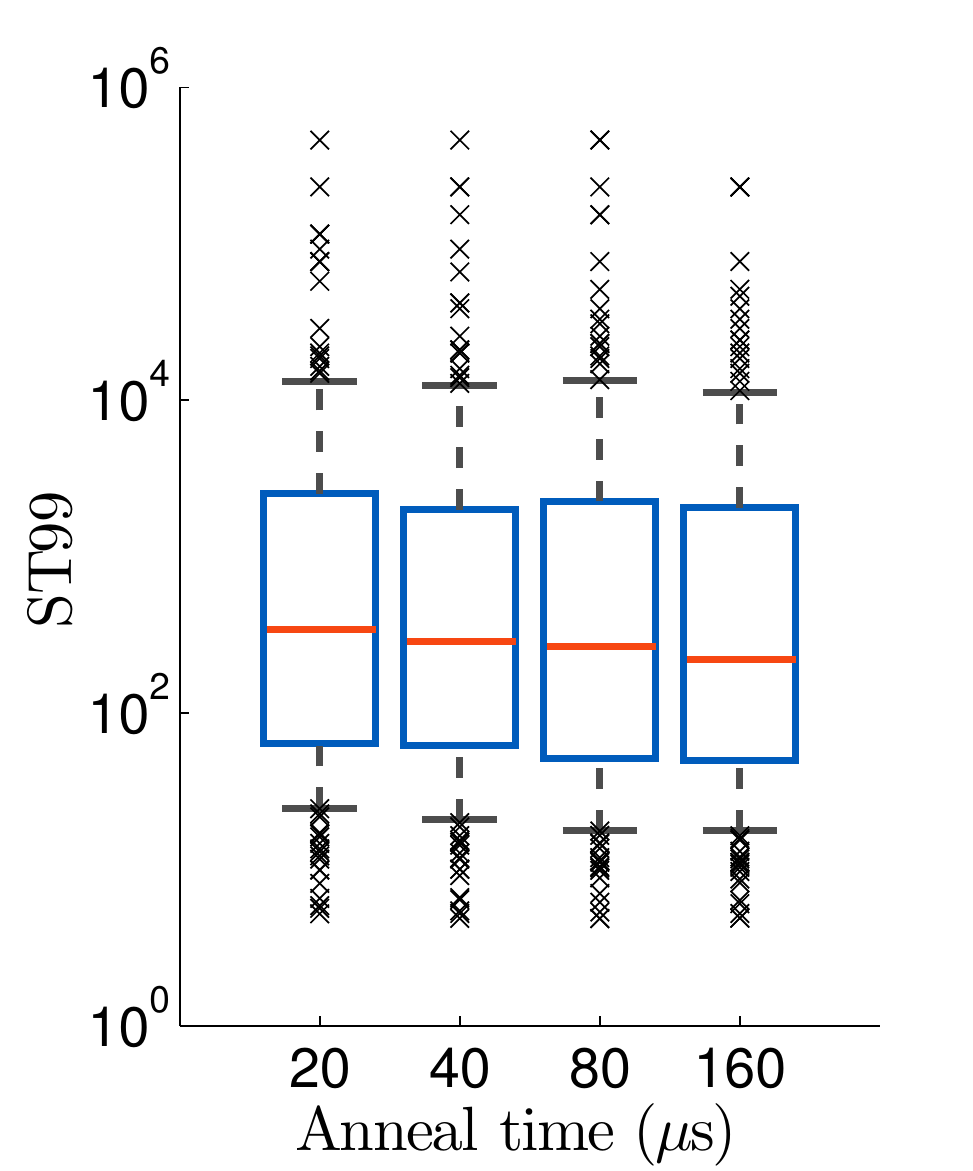}
\includegraphics[height=2.1in]{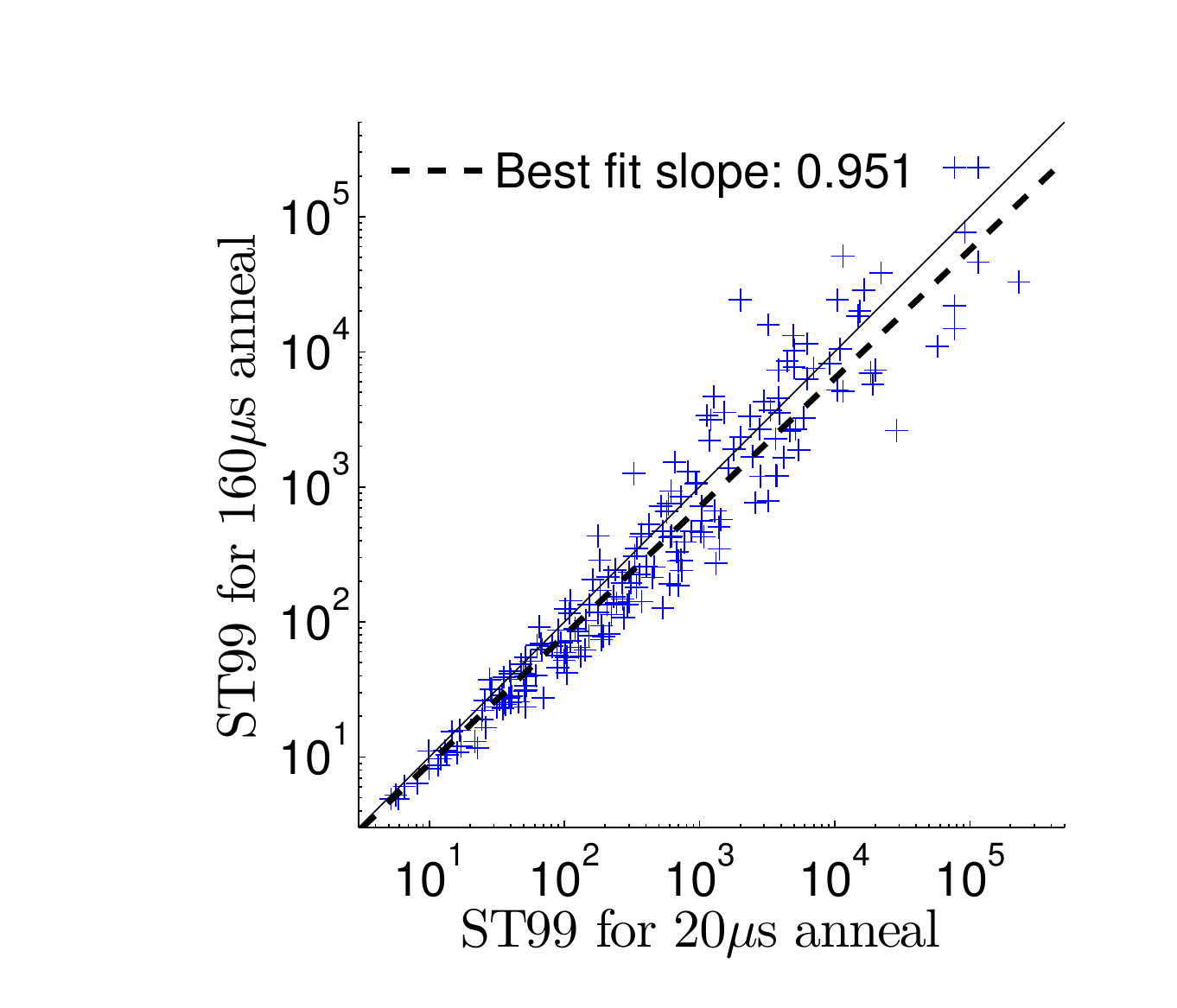}\\%
\includegraphics[height=2.1in]{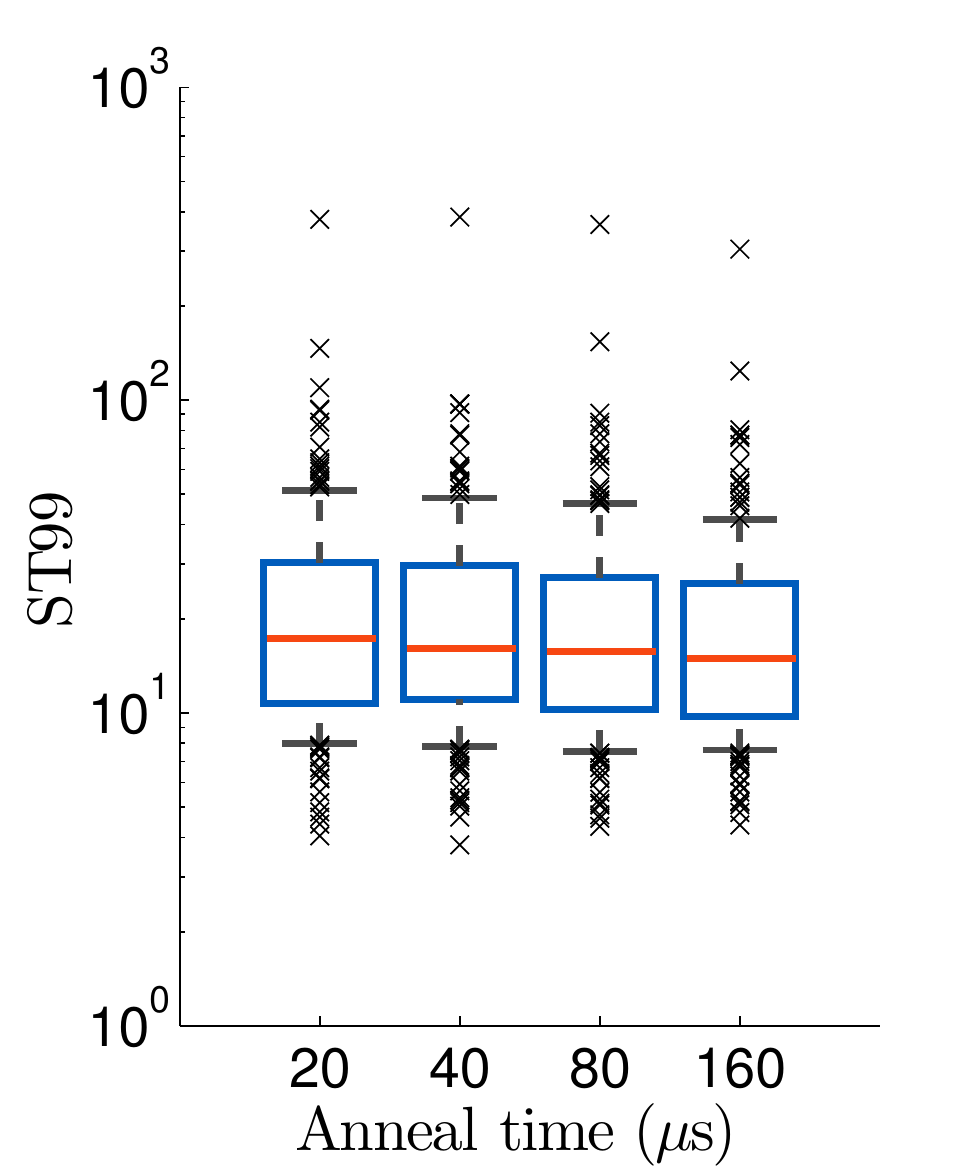}
\includegraphics[height=2.1in]{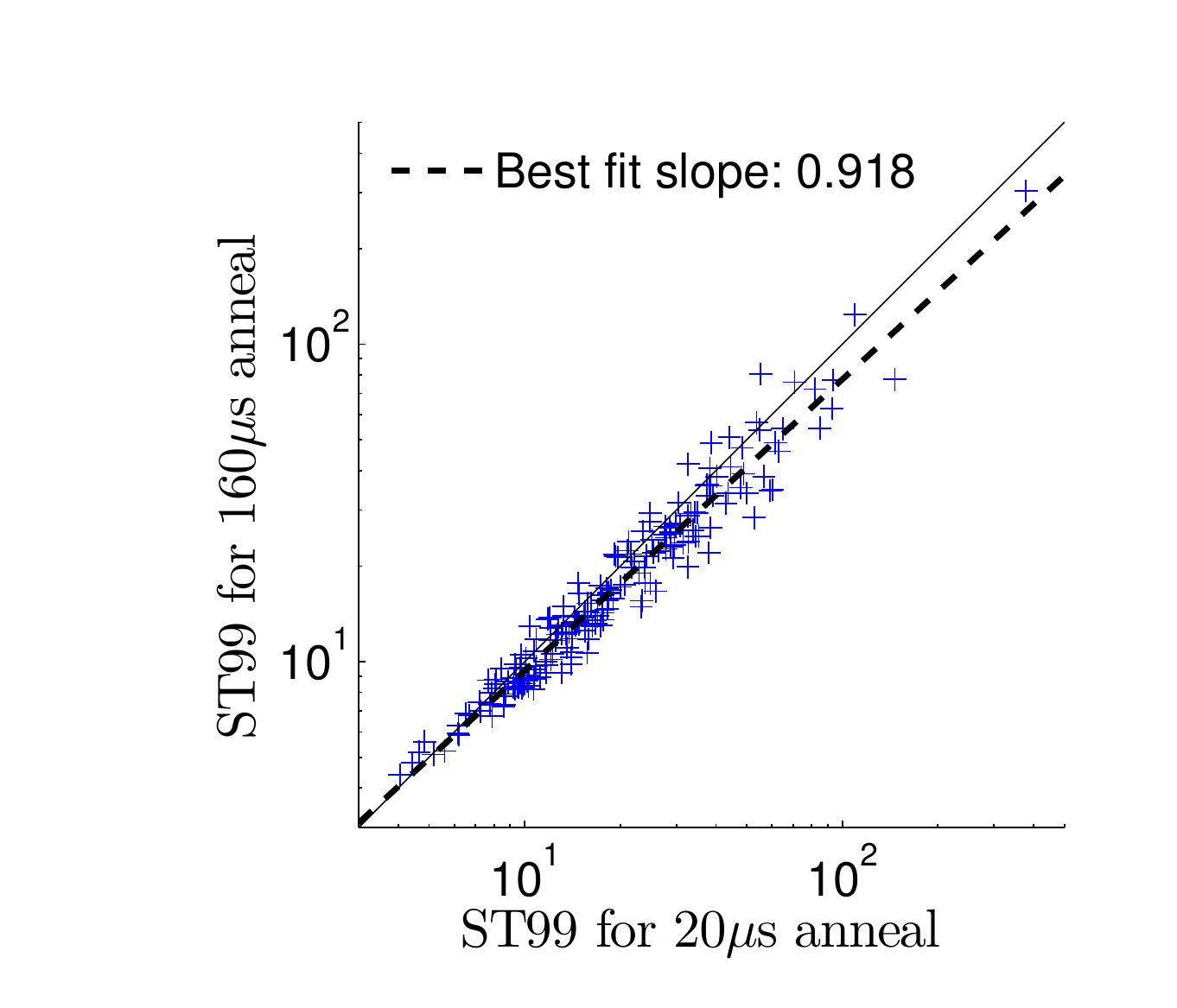}\\%
\end{center}
\caption{\small{Changing anneal times for RAN1 (top) and FL2 (bottom) instances at largest problem size.  Box plots show percentiles 5, 25, 50, 75, and 95, plus outliers,  in 200 instances. }  }
\label{fig.anneal}
\end{figure}

Figure \ref{fig.anneal} shows the result of varying the anneal time from $20\mu$s to $160\mu$s for 200 RAN1 problems and 200 FL2 problems at a 481-qubit scale drawing 100,000 samples over 100 gauge transformations:  despite the noisy data, small reductions in ST99 can be seen at all quantiles.  (Improvements from increased anneal time are less apparent for more error-sensitive classes such as  NAE.)   These limited results  -- together with very preliminary data on a prototype chip with $ >900$ qubits -- suggest that we can expect anneal times to be more important to performance and  to grow above $20\mu$s on next-generation chips with up to 1152 qubits.

\subsection{Methods for minor-embedded problems}\label{sec:minor} 
Suppose we have a Hamiltonian $(h_0,J_0)$ for a general (non-Chimera-structured) IMP instance defined on a graph $G_0$ of $n_0$ vertices; this graph must be minor-embedded in the hardware graph $G=(V,E)$ for solution.  In current D-Wave architectures we have $G \subseteq \mathcal C_k$  where $\mathcal C_k$ is a Chimera graph on $8k^2$ vertices.  Each $\mathcal C_k$ contains $K_{4k}$ as a minor (actually requiring only $4k(k+1)$ qubits \cite{Venturelli2014}),  but in practice we can find more compact embeddings using a heuristic algorithm such as described in \cite{Choi2008}.  (See Appendix A.)

\paragraph{Optimizing chain strength} 
An embedding contains, for each vertex $v_i$ of $(h_0,J_0)$, a set $V_i$ of vertices assigned to a connected subgraph of $G$.  We call each $V_i$ (and a spanning subgraph induced by $V_i$ in $G$) a {\em chain}.  By assigning a strong ferromagnetic coupling (a large-magnitude value $-\kappa$ for $\kappa >0$) between qubits in the same chain we can ensure that in low-energy states of $(h,J)$, all qubits in $V_i$ will take the same spin value, for each $i$.  Thus the hardware output is likely to yield feasible solutions when mapped back to $v_i$ in $(h_0, J_0)$.  

Too-small $\kappa$  produces {\em broken chains}  (i.e.\ chains whose spins do not unanimously agree) in hardware output;    
that is, the  solution in the {\em code space}  cannot be mapped back to the (unembedded) {\em solution space} (see \cite{Young2013}).  
On the other hand, large $\kappa$ decreases the problem scaling factor $\alpha$, which effectively boosts ICE, as in Figure \ref{fig.error-model}.  Therefore the choice of $\kappa$ has a significant effect on hardware success rates.

\begin{figure}
\begin{center}
\includegraphics[height=2.2in]{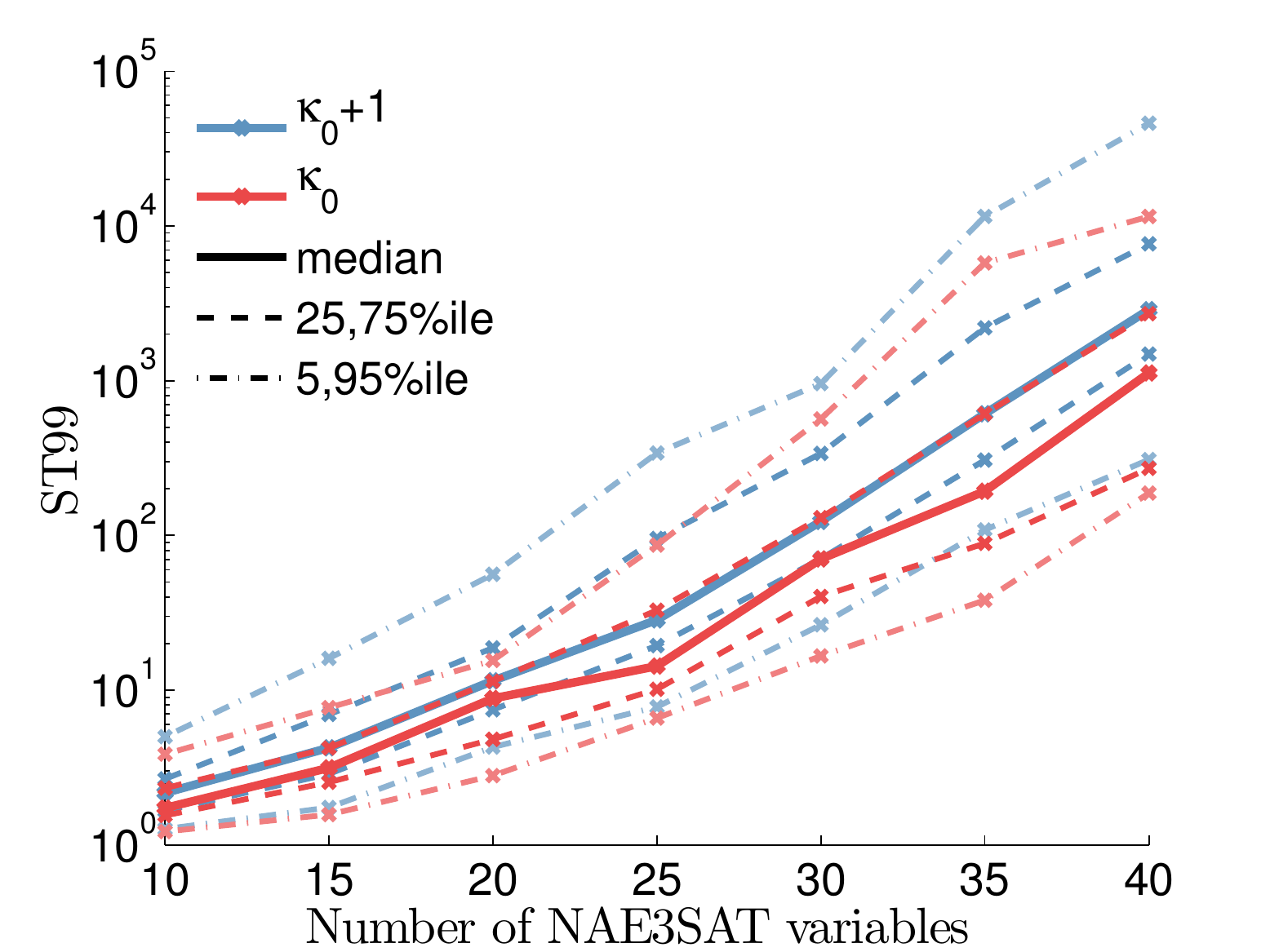}%
\includegraphics[height=2.2in]{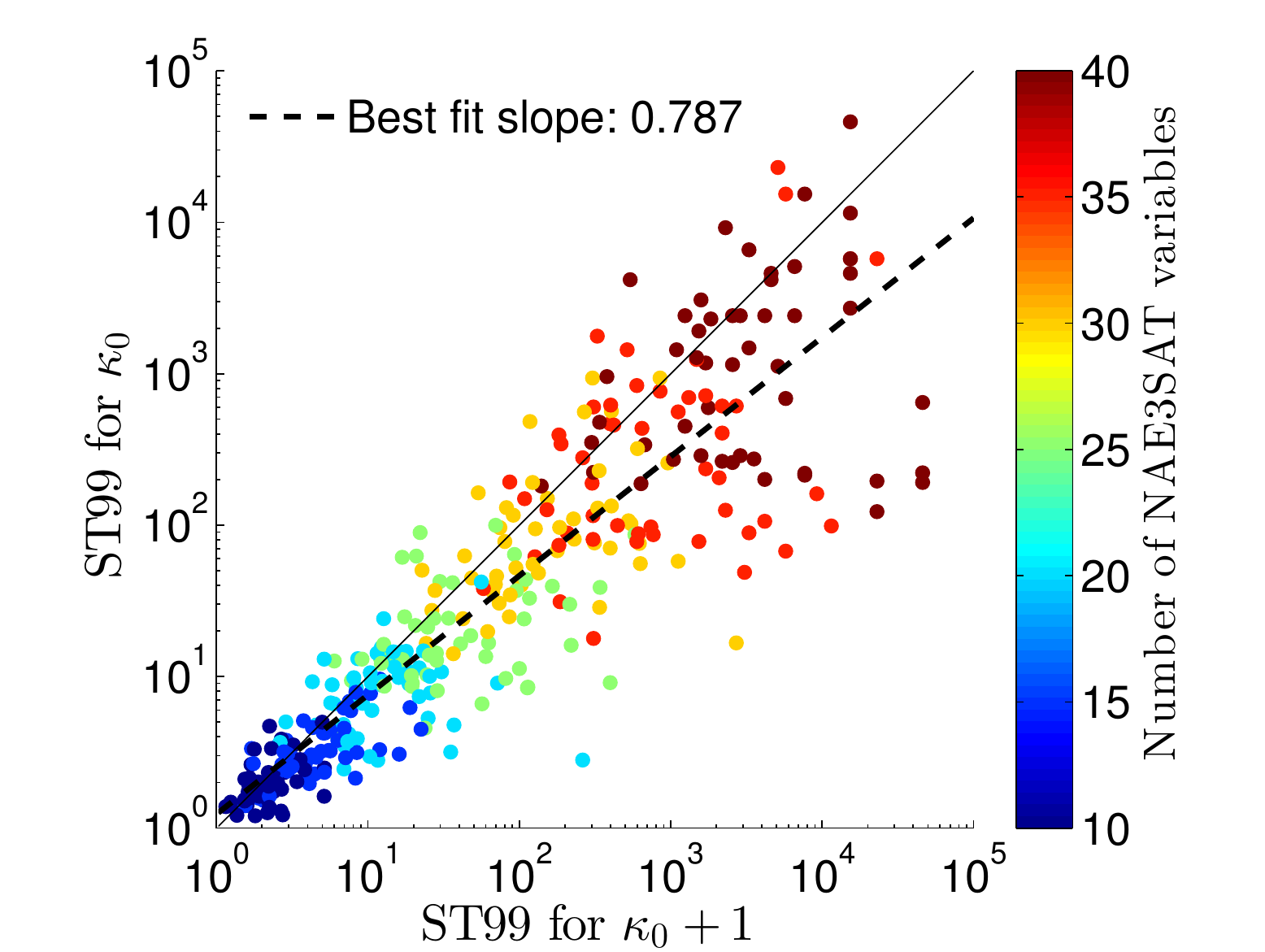}%
\end{center}
\caption{\small{Performance on 50 NAE instances at each problem size using minimal $(\kappa_0)$ and elevated $(\kappa_0+1)$ chain strength for each instance.}}\label{fig.chains}
\end{figure}

For NAE3SAT it appears that the hardware performs best when $\kappa$ is minimized subject to the constraint that no ground state contains a broken chain; results on fully-connected spin glasses appear to agree \cite{Venturelli2014}.  This value of $\kappa$, denoted $\kappa_0$, is instance dependent, and can be approximated empirically by gradually increasing $\kappa$ from zero until the lowest energy found corresponds to a state with no broken chains.  Figure \ref{fig.chains} shows the effect of varying chain strength in NAE instances, with $\kappa\in \{\kappa_0 , \kappa_0+1\}$.  For these instances $\kappa_0$ ranges between 1.5 and 6 on instances of 10 to 40 logical variables, embedded on 18 to 379 physical qubits.   At largest problem sizes, increasing $\kappa_0$ by 1 can more than double median computation times.  The right panel shows a difference of two orders of magnitude on some instances and an interesting bimodal property that awaits further analysis.

\paragraph{Chain shimming}
One can think of the Hamiltonian $(h,J)$ in an embedded problem as a combination of two Hamiltonians, one encoding the original  problem and one encoding the chain constraints.  Thus we have $(h,J) = (h,J_p+J_c)$ since the chain Hamiltonian contains no local fields.  
Due to ICE, $J_c$ introduces a set of effective small local fields called {\em $h$-biases}.  Although ICE will  be mitigated in future hardware generations, this issue can be addressed immediately using a simple technique called {\em chain shimming}.
 
Chain shimming starts by sending the Hamiltonian $(0,J_c)$ to the hardware and measuring the bias on each chain: that is, since $(0,J_c)$ has no local fields and no connections between chains,  the hardware should return unbroken chains having spins $+1$ and $-1$ with equal probability.  If the distribution is biased,  we place a compensating $h$-bias on each qubit of each askew chain.  A few iterations of this process to refine $h$-biases can sometimes improve 
time performance.  This technique can be most efficiently applied when the structure 
of $G_0$ (and therefore the chain Hamiltonian) is constant over many instances,  e.g.\ for the fully-connected graphs described in  \cite{Venturelli2014}.  

\begin{figure}
\begin{center}
\includegraphics[height=2.2in]{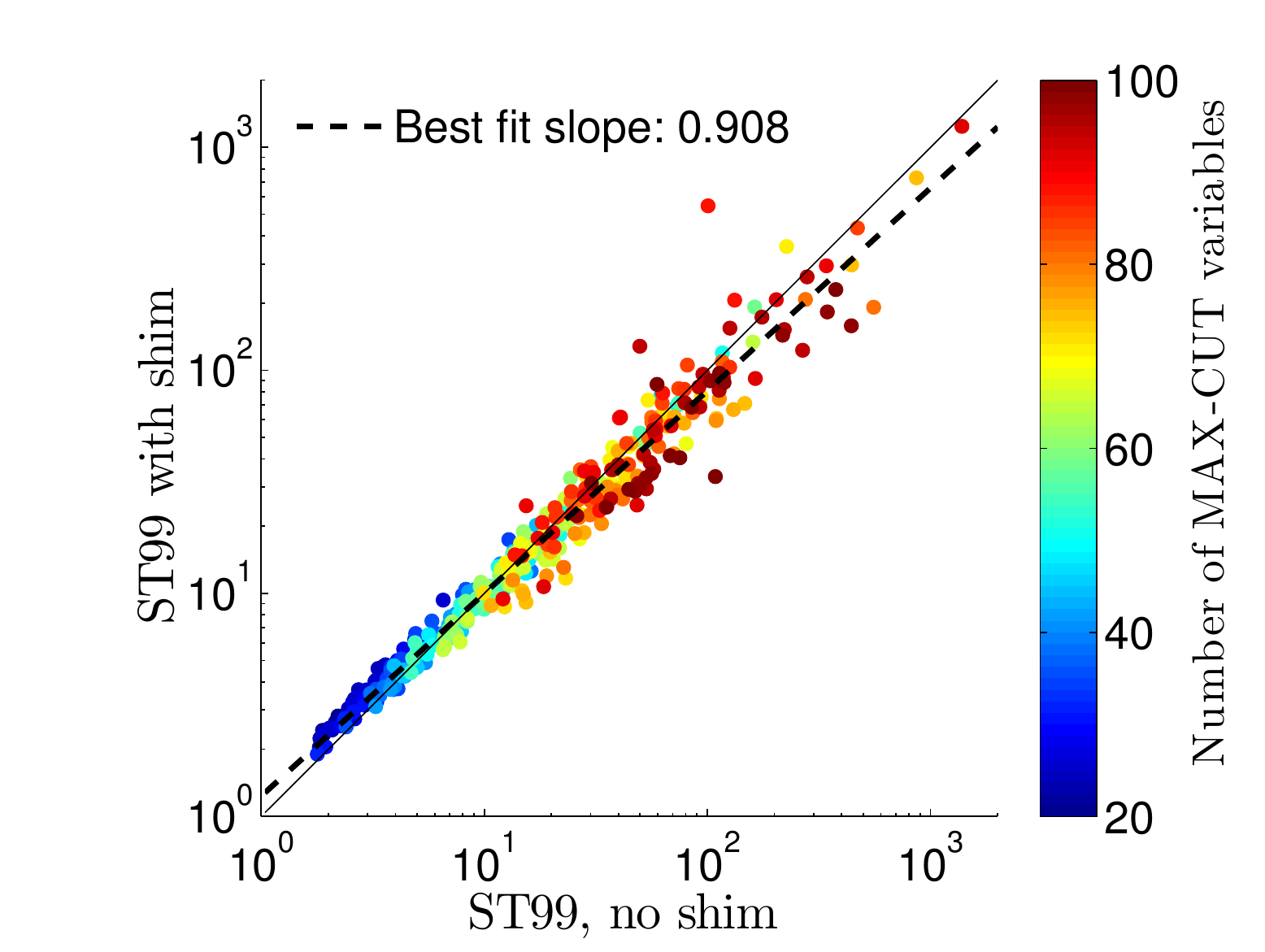}%
\end{center}
\caption{\small{Performance on 10 3MC instances at each even problem size from 20 to 100 logical variables, with and without chain shimming.}}\label{fig.shim} 
\end{figure}

Figure \ref{fig.shim} shows ST99 for 210 3MC instances on V7 with and without shimming.  The data provides some evidence of a slight but systematic improvement in performance as problems become larger and more difficult.  This improvement is not seen for NAE instances, likely due to the higher chain strength required for NAE instances and the subsequent ICE sensitivity (3MC instances use $\kappa=2$, which is always sufficient).

\subsection{Classical Error Correction via Postprocessing}
An obvious remedy for some  types of errors described in \ref{sec:error} is to apply error correction techniques.   
Pudenz {\em et al.}\ \cite{Pudenz2014,Pudenz2014a} present {\em quantum error-correcting codes} for D-Wave architectures; and Young {\em et al.}\ \cite{Young2013} describe {\em quantum stabilizer codes}.  These techniques boost hardware performance immensely at the cost of many ancillary and redundant qubits, and consequently a reduction in the size of problems  that can be solved on a fixed-size chip.   An alternative strategy discussed here is to apply cheap classical postprocessing operations to the solutions returned by hardware.

\paragraph{Majority vote}  In embedded problems it is possible for the hardware to return solutions with broken chains.  Rather than discarding such samples, we may instead set the spin of each qubit in a chain according to a {\em majority vote} of qubits in the same chain (breaking ties randomly).  This is computationally inexpensive and improves hardware success probabilities.  Several more sophisticated methods may be considered for repairing broken chains, such as increasing chain strength until votes are unanimous or converting unanimous chains into local fields to reduce the problem:  further study is needed.    

\paragraph{Greedy descent}  Another simple postprocessing technique is to walk each hardware solution down to a local minimum by repeatedly flipping random bits to strictly reduce solution cost.  
We call this approach {\em greedy descent}.   In a minor-embedded problem, this can be applied to the solution to the unembedded or the  embedded problem, or both.  More generally, one can apply as a postprocessing step any classical heuristic that takes an initial state from the hardware and refines it, e.g.\ simulated annealing,  tabu search, or parallel tempering.\footnote{We recognize that there is conceptually a fine line between using a DW2 system as a  preprocessor and using classical heuristic as an error-correcting postprocessor.  Work is underway to explore these ideas.} 

\begin{figure}
\begin{center}
\includegraphics[height=2.2in]{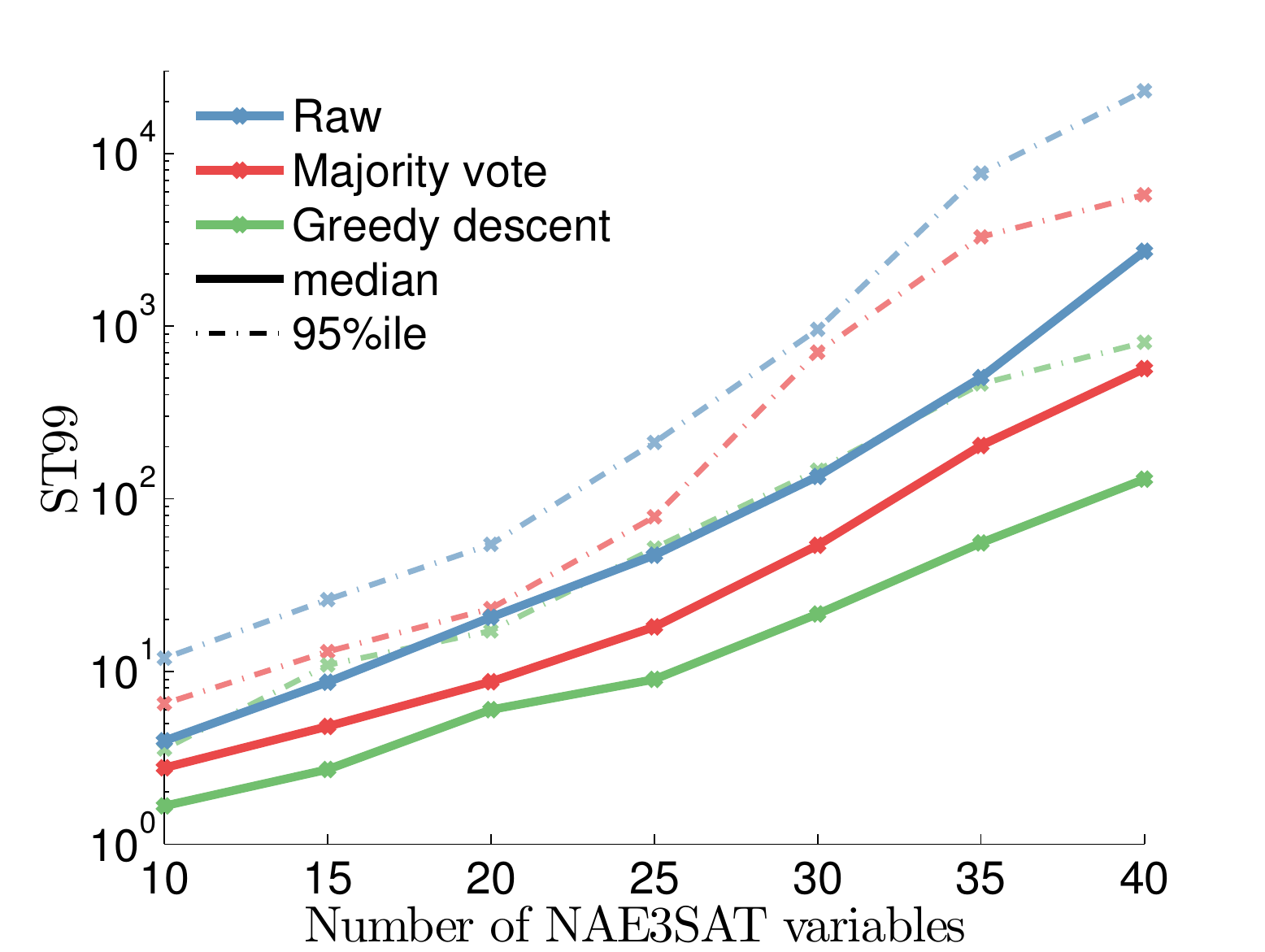}%
\end{center}
\caption{\small{Error correction by postprocessing on 50 NAE problems on each problem size.}}\label{fig.post} 
\end{figure}

Figure \ref{fig.post} shows the effect of postprocessing on NAE instances.  In these tests $\kappa$ is set to $\kappa_0+1$, runs at higher $\kappa$ will derive less benefit from majority vote, and more from greedy descent.

\section{Conclusions}
We have presented several algorithm engineering techniques that aim to improve the performance of D-Wave quantum annealing processors.   These include strategies for modifying anneal times, changing the problem Hamiltonian (gauge transformations, chain shimming),  improving chains in embedded problems,  and exploiting simple postprocessing ideas.   Many more ideas along these lines can be identified, and it remains to be seen what performance gains can be achieved by applying combinations of techniques.   Beyond these individual strategies, perhaps a more important contribution has been the presentation of a conceptual framework for distinguishing performance of the quantum algorithm from its realization in technologically immature but rapidly-developing hardware.      

The question arises as to how some of these techniques might affect the performance of classical software solvers.  Techniques that focus on mitigating Hamiltonian misspecification (e.g.\ chain shimming and  gauge transformation) are largely irrelevant to classical heuristic approaches to solving IMP,  since digital computers do not experience these types of errors.  
Other techniques such as postprocessing and longer anneal times can be successfully transferred to some algorithmic approaches -- such as heuristic search --  but not necessarily to others -- such as dynamic programming based approaches.

Both the quantum annealing paradigm and its implementation on quantum hardware are very new concepts, 
and the current performance model  is primitive and incomplete.   
This paper represents a small step towards better understanding of performance in this novel computing paradigm.    

\section{Acknowledgements}
We extend warm thanks for useful discussions and suggestions to Carrie Cheung, Brandon Denis, Itay Hen, Robert Israel, Jamie King, Trevor Lanting, Aidan Roy, Miles Steininger, Murray Thom, and Cong Wang. 

\bibliography{bibtex}{}

\appendix
\section{Chimera structure and the hardware graph}

\begin{figure}
\begin{center}\includegraphics[width=1.5in]{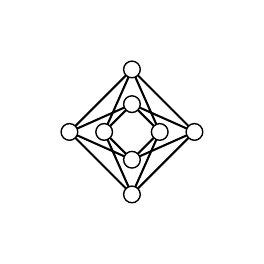}\end{center}
\caption{\small{A single Chimera cell, $K_{4,4}$.}}
\label{app:chimera}
\end{figure} 

\begin{figure}
\begin{center}\includegraphics[width=4.1in]{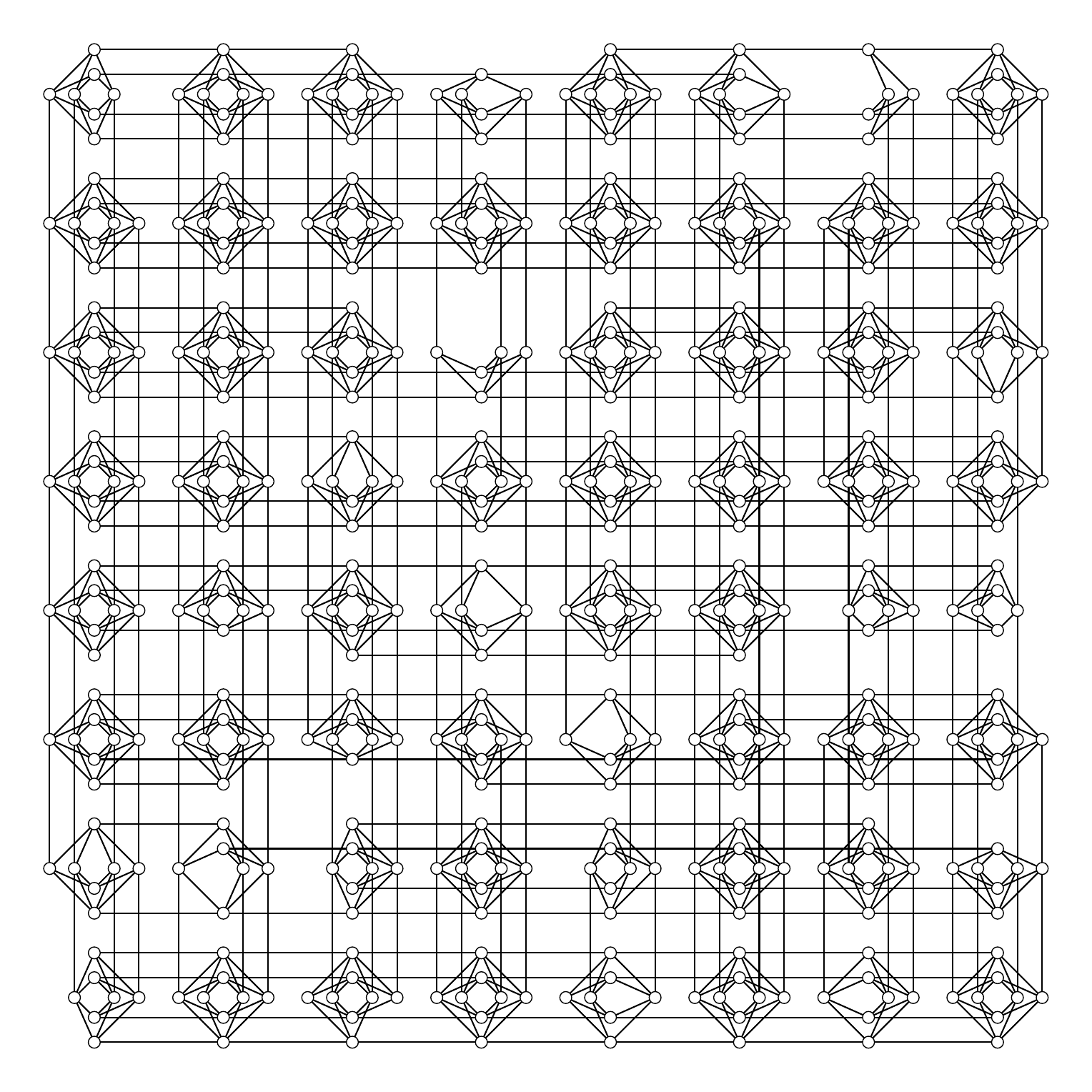}\end{center}
\caption{\small{The V7 hardware graph, a subgraph of $\mathcal C_8$.} }
\label{app:v7} 
\end{figure}

A {\em Chimera graph} $\mathcal C_k$ consists of a $k \times k$ grid of {\em cells}.  In current D-Wave configurations each cell is a complete bipartite graph $K_{4,4}$.  Vertices in a row are matched to corresponding vertices in neighboring cells above and below, and vertices in a column are matched to corresponding vertices in neighbouring cells to the left and right.  See Figures \ref{app:chimera} and \ref{app:v7}.   A $\mathcal C_k$ contains $8k^2$  vertices of degree 6 (internal vertices), and 5 (sides),  totalling  $24k^2 - 16k$ edges.  The hardware graph of V7 is a subgraph of $\mathcal C_8$, a result of fabrication imperfections and high calibration throughput.  The working graph varies from chip to chip.

\paragraph{Minor-embeddings}
A {\em  minor} of a given graph $G = (V,E)$ is any graph that can be constructed from $G$ by application of some number of the following operations,  in any order:
\begin{enumerate}
\item Remove an edge.
\item Remove a vertex and incident edges.
\item Contract an edge,  combining its incident vertices.   
\end{enumerate} 
 
If graph $G'$ is a minor of  graph $G$, it is straightforward to reduce IMP on $G'$ to IMP on $G$:  for each edge $(uv)$ of $G$ that is contracted in the graph minor construction, assign to $J_{uv}$ a strong ferromagnetic (negative) coupling.  If $J_{uv}$ is sufficiently large, $u$ and $v$ will take the same spin in any low-energy configuration.  As discussed in the paper,  sufficient bounds on $J_{uv}$ are highly dependent on the structure of the individual minor chosen.  

Choi \cite{Choi2008} shows that a complete graph of  $n = 4k$ vertices can be minor-embedded in the upper diagonal of a $\mathcal C_k$, using $4k(k+1)$ vertices.  The problem complexity of deciding the minor-embeddability of an arbitrary graph into a Chimera graph is open.  

\section{Instance classes}\label{appendix:testbed}
We present a brief overview of instance classes used in this work. 

\subsection{Random instances (RAN$R$)} 
For given hardware graph $H = (V,E)$,  for each $(u,v) \in E$ generate a weight uniformly at random from the integer range $[-R \ldots +R]$ 
(omitting 0). 

Katzgraber et al \cite{Katzgraber2014} have shown that these instances are fairly easy for simulated annealing based 
solvers,  for two reasons.   First, a random instance typically has a large number of global minima, which can be found using many random restarts.   Second, during most of the anneal time the solution landscape has gentle slopes and no high barriers: thus the correct neighborhood of a global optimum is found early in the anneal 
process.    

\subsection{Frustrated loop instances}

We present a construction of Hen \cite{henaqc2014} for an Ising Hamiltonian $(h,J)$ over a hardware graph $G$ in which $\uparrow\uparrow\ldots\uparrow$ is a ground state.  Let $R$, our precision limit, be a positive integer, let $n$ be the number of vertices in $G$, and let $r$ be a constraint-to-qubit ratio.  We construct a Hamiltonian consisting of a conjunction of $[rn]$ frustrated loops (where $[rn]$ denotes the round-off of $rn$) as follows.

First let $c_1$ be a cycle chosen at random in some way.  Here we do this by performing a random walk in $G$ starting at a random vertex, and taking the first cycle we find.  To ensure that the cycles spread across $G$ sufficiently, we reject a cycle if it is contained entirely in a Chimera $K_{4,4}$ unit cell, and repeat the construction.  Let $n_1$ be the number of vertices in $c_1$; note that due to the structure of $G$, $n_1$ is even and at least $6$.  We construct a Hamiltonian $(0,J^{(1)})$ by setting every edge of $c_1$ to $-1$ except one chosen uniformly at random, which we set to $1$.  It is now straightforward to check that $(0,J^{(1)})$ has $2n_1$ ground states, and ground state energy $-(n-2)$.

We repeat this construction for further cycles $c_2,\ldots,c_{rn}$, with the following wrinkle: if after choosing $k$ cycles, an edge $uv$ of $G$ has
\[
\left| \sum_{i=1}^k J_{uv}^{(i)}\right| = R,
\]
we forbid the edge $uv$ from appearing in cycles $c_{k+1},\ldots,c_{rn}$.

The final Hamiltonian of the problem is $(h,J)=(0,\sum_{i=1}^{[rn]}J^{(i)})$.  Note that the specified ground state $\uparrow\uparrow\ldots\uparrow$ can be ``hidden'' by applying a gauge transformation to the Hamiltonian.

The instances we use in this paper have ratio $r=0.2$, which roughly corresponds to an empirically observed phase transition \cite{henaqc2014}, and precision limit $R=2$, which is the minimum possible value that allows a rich set of instances.

\subsection{Random cubic MAX-CUT instances}

MAX-CUT on cubic graphs is a well-known NP-hard problem \cite{berman1999some} that has a very simple Ising formulation.  Maximum cardinality cuts on a graph $G$ correspond to ground states of the Ising problem $(0,J)$ where $J_{uv}=1$ for all $uv\in E(G)$, and $J_{uv}=0$ elsewhere.  It is straightforward to confirm that in the cubic case, when embedding a MAX-CUT Hamiltonian, chain strength $\kappa=2$ is always sufficient to guarantee chain fidelity in the ground state.  Indeed, any $\kappa>1$ is sufficient.

\subsection{NAE3SAT instances}

As in FL$R$ instances described above, we construct NAE instances as the conjunction of $[rn]$ constraints.  In this case we use $r=2.1$, which corresponds roughly to the phase transition for Not-all-equal 3-SAT \cite{achlioptas2001phase}.  Further, the instances must be minor-embedded, as they do not naturally fit into the Chimera hardware graph.

We generate a random NAE3SAT instance by choosing $[rn]$ clauses at random.  Each clause consists of 3 unique randomly selected variables, each of which is negated independently with probability $\tfrac 12$.  The Hamiltonian for a clause $(q_1x_1,q_2x_2,q_3x_3)$, where $q_i=-1$ if $x_i$ is negated in the clause and $q_i=1$ otherwise, has $h=0$, and all entries of $J$ zero except $J^{(i)}_{x_i,x_j} = q_iq_j$ for $1\leq i<j\leq 3$.  As with frustrated loops, our final Hamiltonian $(h,J)$ has $h=0$ and $J=\sum_{i=1}^{[rn]}J^{(i)}$.

For sufficiently large $n$, the adjacency graph of the nonzero entries of $J$ is sparse, with average degree $6r$, and the nonzero entries of $J$ are overwhelmingly in $\{-1,1\}$.  These random instances are converted to Chimera-structured problems via the heuristic minor-embedding algorithm described in \cite{Cai2014}.  The question of how performance varies from one embedding of an instance to another requires further study outside the scope of this paper.  To separate this issue from the algorithm engineering approaches we study here, we take five embeddings of each instance.  When we want to compare the performance under several parameter settings, we choose the ``best'' embedding to study.  That is, we choose the embedding for each instance that maximizes the geometric mean of $\pi$ under the parameter settings we compare.

  Our choice of $\kappa_0$ for each embedded instance was (over)estimated by solving each problem for chain strength in $1,1.5,2,\ldots$ until a ground state without broken chains was found.

\end{document}